\documentclass[aps,prl,groupedaddress,preprint,showpacs]{revtex4}
\usepackage{graphicx} 
\begin{document}
\preprint{UPoN-2012, February 20-24, Kalkata, India}
\def\Journal#1#2#3#4{{\it #1} {\bf #2} (#3) {#4}} 
 
\def\BiJ{ Biophys. J.}                 
\def\Bios{ Biosensors and Bioelectronics} 
\def\LNC{ Lett. Nuovo Cimento} 
\def\JCP{ J. Chem. Phys.} 
\def\JAP{ J. Appl. Phys.} 
\def\JMB{ J. Mol. Biol.} 
\def\CMP{ Comm. Math. Phys.} 
\def\LMP{ Lett. Math. Phys.} 
\def\NLE{{ Nature Lett.}} 
\def\NPB{{ Nucl. Phys.} B} 
\def\PLA{{ Phys. Lett.}  A} 
\def\PLB{{ Phys. Lett.}  B} 
\def\PRL{ Phys. Rev. Lett.} 
\def\PRA{{ Phys. Rev.} A} 
\def\PRE{{ Phys. Rev.} E} 
\def\PRB{{ Phys. Rev.} B} 
\def\EPL{{Europhys. Lett.} } 
\def\PD{{ Physica} D} 
\def\ZPC{{ Z. Phys.} C} 
\def\RMP{ Rev. Mod. Phys.} 
\def\EPJD{{ Eur. Phys. J.} D} 
\def\SAB{ Sens. Act. B} 
\title{Gumbel distribution and current fluctuations in critical systems}

\author{E. ALFINITO}
\address{Dipartimento di Ingegneria dell'Innovazione, Universit\`a del Salento, via Monteroni\\
Lecce, Italy, \\ CNISM (Consorzio Interuniversitario per le Scienze Fisiche della Materia)
\footnote{eleonora.alfinito@unisalento.it}}

\author{J.-F. MILLITHALER   }
\address{Dipartimento di Ingegneria dell'Innovazione, Universit\`a del Salento, via Monteroni\\
Lecce, Italy, \\ CNISM (Consorzio Interuniversitario per le Scienze Fisiche della Materia)
\footnote{jf.millithaler@unisalento.it}
}

\author{L. REGGIANI}

\address{Dipartimento di Matematica e
Fisica, "Ennio De Giorgi", Universit\`a del Salento, via Monteroni, 73100
Lecce, Italy \\ CNISM (Consorzio Interuniversitario per le Scienze Fisiche della Materia)
\footnote{lino.reggiani@unisalento.it}
}

\begin{abstract}
We investigate a particular phase transition between two different tunneling regimes, direct  and injection (Fowler-Nordheim),
experimentally observed in the current-voltage characteristics of
the light receptor bacteriorhodopsin (bR).
Here, the sharp increase of the current above about 3 V is theoretically interpreted as the cross-over between the direct and injection sequential-tunneling regimes.
Theory also predicts a very special behaviour for the associated current fluctuations around steady state.
We find the remarkable result that in a large range of bias around the transition
between the two tunneling regimes,  the  probability density functions can be traced back to the generalization of the Gumbel distribution. This non-Gaussian distribution is the universal standard  to describe fluctuations under extreme conditions. 
\end{abstract}

\maketitle
\section{Introduction}
Non Gaussian  distributions (NGDs) evidence deviations from the central limit theorem, ubiquitous under thermal equilibrium conditions.
Among the NGDs, increasing attention has been recently devoted to the case of the generalized Gumbel distribution $G(a)$, related to the statistics of extreme events.
Since the pioneer papers \cite{Gumbel,Bramwell,Noullez}, Gumbel-like distributions have been evidenced in a wide series of extreme events, ranging from physical or sociological and environment scenarios \cite{Aji,Bertin,Brey,Clusel,Ciliberto,Manzato}.
The possibility that the generalized Gumbel distribution can be associated with the identification of some universal behaviour of a wide class of extreme events is emerging as an intriguing issue \cite{Noullez,Bertin,Clusel,Ciliberto}.
\par
The aim of this paper is to present evidence of the $G(1)$ distribution in the current fluctuations of a monolayer of bacteriorhodopsin (bR), a protein
present in the Archean organism \textit{Halobacterium salinarum}\cite{Corcelli} and acting as a light
receptor.
\par
In the framework of a microscopic interpretation of electrical transport through this protein, a model, called INPA (impedance network protein analogue)\cite{PRE11,Epl}, reproduces the current-voltage (I-V) characteristics in agreement with experiments
\cite{Gomila};
on the other hand, it also produces the current fluctuations on which no data are presently available.
\begin{figure}[htb]
\centering\noindent
\includegraphics[scale=0.7]{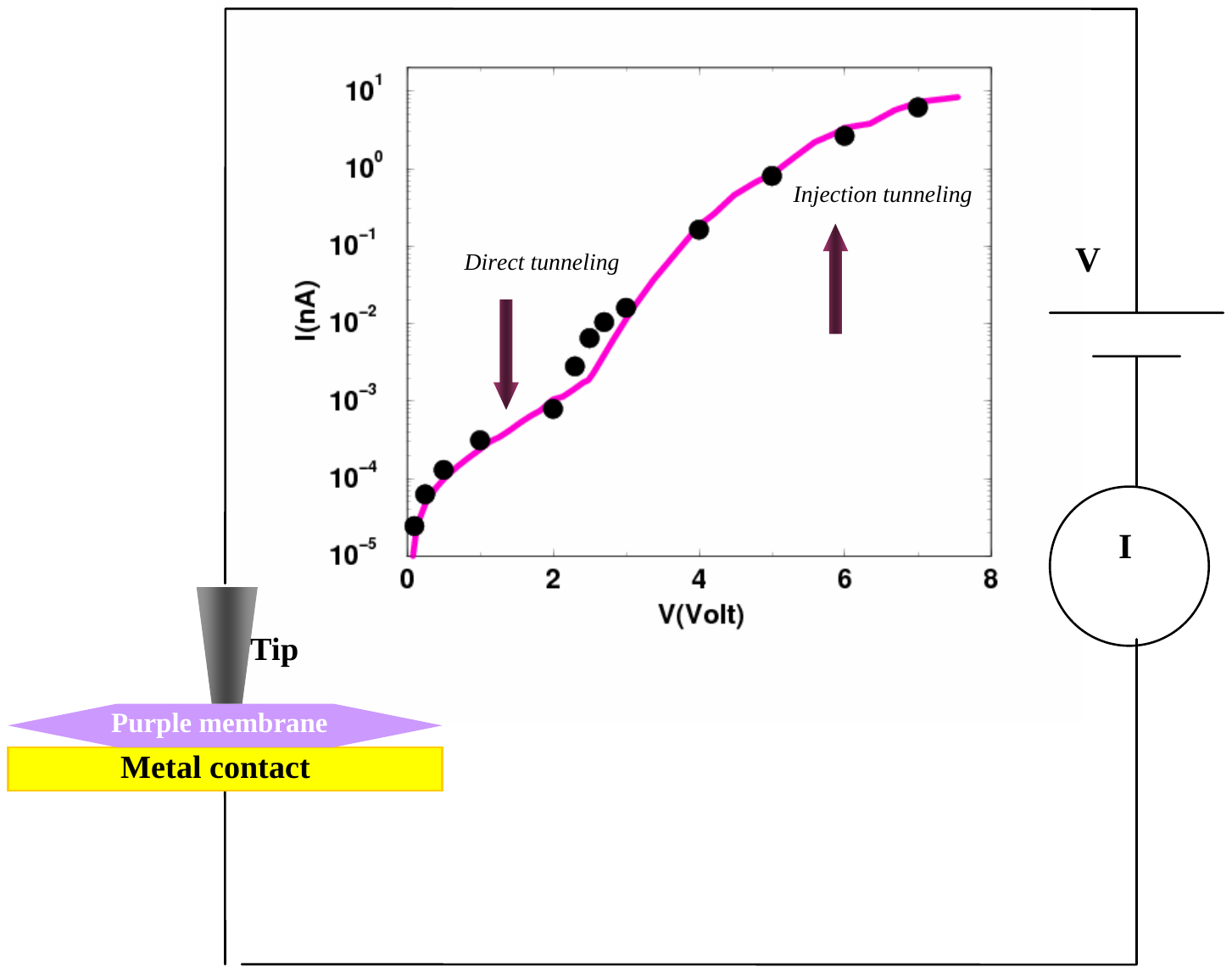}
\caption{A schematic illustration of the experimental apparatus used for the I-V characterization. The experiments (continuous curve) and the theoretical results (full circles) are reported in the center of the figure.}
\label{fig:1}
\end{figure}
%
A set of I-V characteristics were carried out on bR monolayers, making use of an atomic force microscope (AFM) technique. Measurements were performed at the nano-scale length and  evidenced a cross-over between a direct (DT) and an injection or Fowler-Nordheim (FN) regime of the associated tunneling current \cite{Gomila}.
The aforementioned INPA succeeded in providing a microscopic interpretation of experiments, by using an interpolation scheme that accounts for a continuous transition between the two tunneling regimes.  The mechanism of charge transfer among the amino-acids constituting the tertiary structure of the protein, is taken to be a sequential tunneling.
\par
Figure \ref{fig:1} shows a sketch of the experimental apparatus. The I-V characteristics, as obtained by the experiments (see continuous curve) and the theoretical model (see the full circles) when the AFM tip just touches the protein films at about 4.6 nm from the bottom metal-contact, is also reported.    
The transport approach uses the Monte Carlo technique and  allows for the simultaneous detection of the current and its fluctuations around the steady state.

Interestingly enough, we found that the cross-over between DT and FN regimes is much strongly marked by the variance of current fluctuations than by the current variation itself.
Moreover, the associated NGDs are found to follow a generalized Gumbel distribution\cite{Noullez,Bertin}.
\par
The content of the paper is the  investigation of the properties of the current fluctuations and is organized as follows.
In the next section we will briefly recall the microscopic model used to interpret the experiments \cite{Gomila}; this model  predicts NGDs for the current fluctuations around  steady state.
Section 3 will briefly summarize the properties of the \textit{scaled} Gumbel distribution, $G(1)$.
Section 4 will report the results of the theoretical predictions and discuss the
implications of the NGDs of current fluctuations exhibited by the simulations.
Major conclusions will be reported in Sec. 5. 

\section{Theory}
\begin{figure}[htb]
\centering\noindent
\includegraphics[scale=0.7]{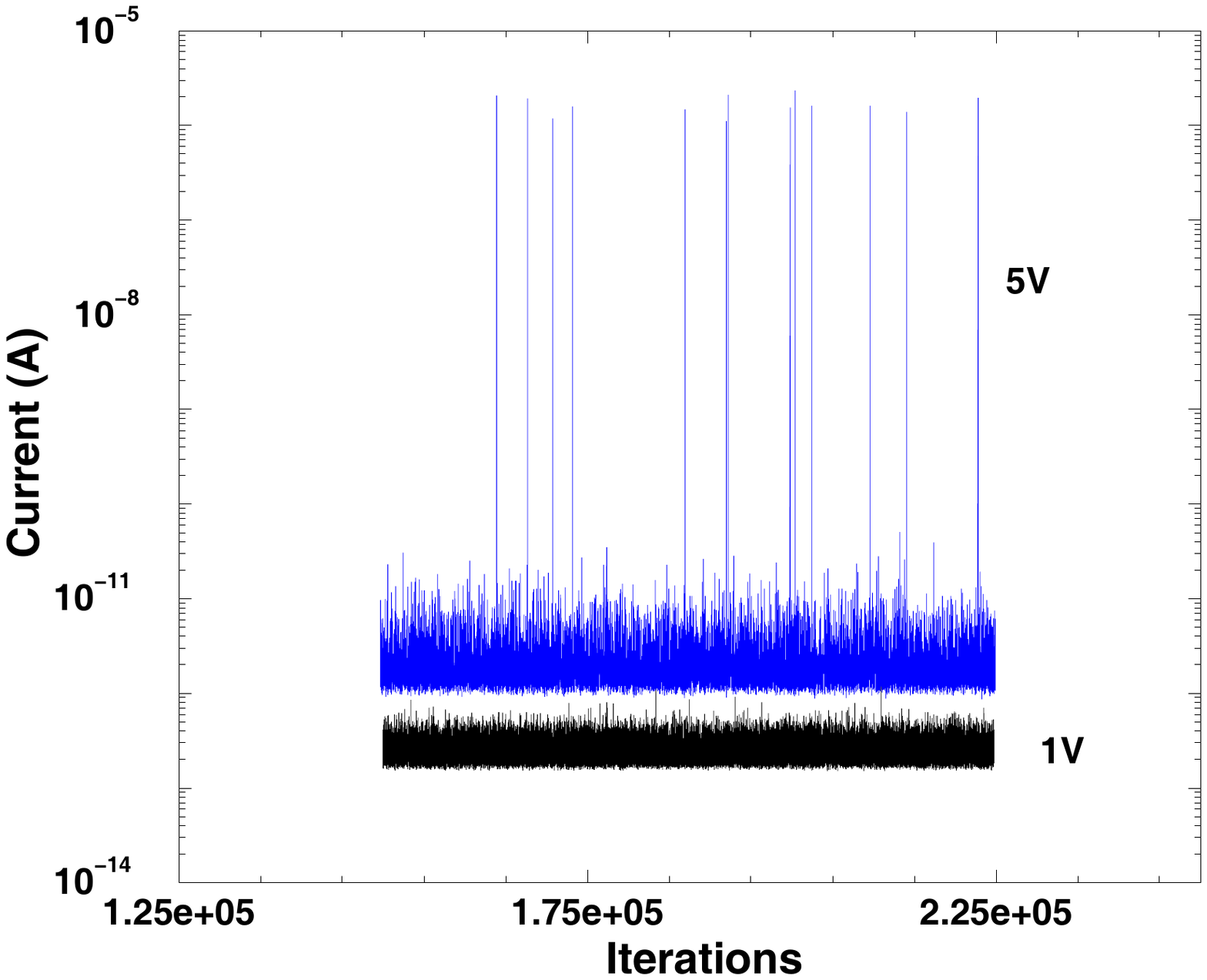}
\caption{Evolution (in units of iterations) of the instantaneous current responses for a low bias of 1 V, when direct tunneling regime occurs, and for a high bias of 5 V, when injection tunneling regime occurs.}
\label{fig:2}
\end{figure}
%
The layout of the microscopic INPA approach is briefly summarized in the following.
\begin{figure}[htb]
\centering\noindent
\includegraphics[scale=0.7]{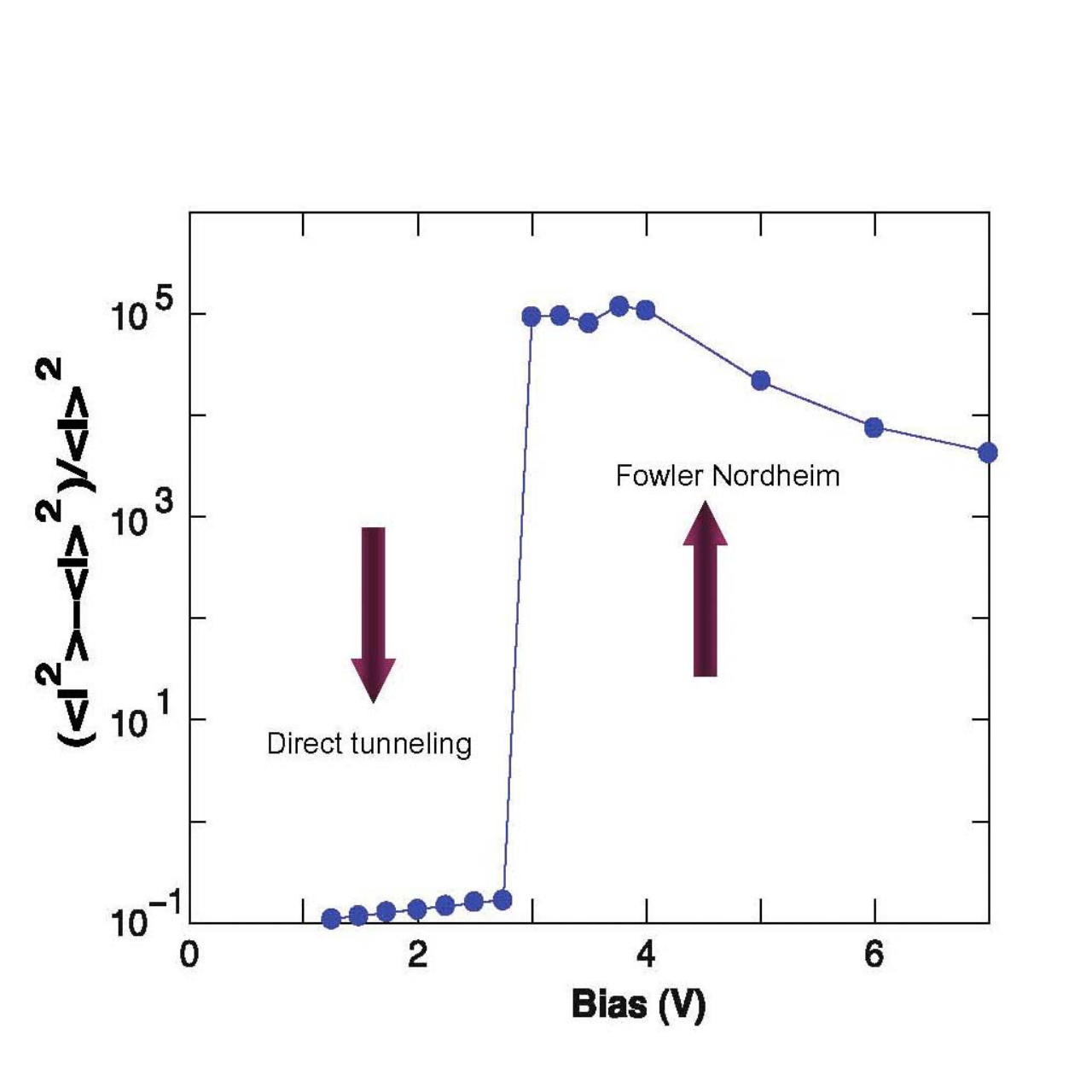}
\caption{Variance of current fluctuations normalized to the value of the steady state current vs applied voltage. The continuous curve is a guide to the eyes.}
\label{fig:3}
\end{figure}
%
The single protein is mapped into an impedance network, by using the C$_\alpha$ atom of each amino acid as a node and by introducing an elemental impedance (in general an RC circuit) as a link between neighbouring nodes. 
The number of links starting from each node is determined by a
cut-off interaction radius, say $R_{C}$, and by the network topology (not
regular).
In the present analysis, we choose $R_{C}$ = 6 \AA, a value that optimizes the native to activated state resolution of the protein \cite{Epl}. 
The value of the elemental impedance (a simple resistance in this case) depends on the distance between amino-acids as:
$r_{i,j}=\rho\, l_{i,j}/\mathcal{A}_{i,j}$, 
where $\rho$ is the resistivity. The value of resistivity is here taken to be the same for all the links, and depends on the voltage drop between the corresponding nodes, as detailed below. The pedices $i,j$ refer to the amino-acids between which the link is stretched, $l_{i,j}$ is the distance between the labeled amino-acids, taken as point-like centers,  and ${\mathcal{A}_{i,j}}$ is the cross-sectional area shared by the labeled amino-acids:
${\mathcal{A}_{i,j}}= \pi\left(\textsl{ R}_C^{2}-l^{2}_{i,j}/4\right)$ \cite{Nano}.
\par
To take into account the superlinear features of the I-V characteristics
\cite{Gomila}, the link resistivity, $\rho$,  is chosen to depend on the voltage drop between nodes as:
\begin{equation}
\rho(V)=\left\{\begin{array}{lll}
\rho_{MAX}& \hspace{.5cm }& eV <  \Phi  \\ \\
 \rho_{MAX} (\frac{\Phi}{eV})+\rho_{min}(1- \frac{\Phi}{eV}) &\hspace{.5cm} & eV \ge  \Phi 
 \end{array}
  \right.
\label{eq:3}
\end{equation}
where $\rho_{MAX}$ is the resistivity value used to fit the I-V characteristic at the lowest voltages,  $\rho_{min} \ll \rho_{MAX}$  is an extremely low series resistance limiting the current at the highest voltages,  and $\Phi$  is the height of the tunneling barrier between nodes.
Since charge transfer is here interpreted in terms of a sequential tunneling between neighbouring amino-acids, the above interpolation formula reflects the different voltage dependence in the prefactor of the current expression \cite{Wang}: $I\sim V$ in the DT regime, and $I\sim V^2$ in the FN tunneling regime.
\par
For the transmission probability of the tunneling mechanism we take the expression given by Ref. \cite{Simmons}.
The microscopic model of electron transport used for describing the experimental data, is based on the local possibility of choosing between different tunneling probabilities given by:
\begin{equation} 
\mathcal{P}^{D}_{i,j}= \exp \left[- \frac{2 l_{i,j}}{\hbar} \sqrt{2m(\Phi-\frac{1}{2}
eV_{i,j})} \right] \ ,
\hspace{0.7cm}
 eV_{i,j}  < \Phi  \,
\label{eq:1}
\end{equation}
\begin{equation}\label{eq:2}
\mathcal{P}^{FN}_{ij}=\exp \left[-\left(\frac{2l_{i,j}\sqrt{2m}}{\hbar}\right)\frac{\Phi}{eV_{i,j}}\sqrt{\frac{\Phi}{2}} \right] \ , 
\hspace{0.7cm}
 eV_{i,j} \ge \Phi  \ ;
\end{equation}
where $V_{i,j}$ is the potential drop between the couple of $i,j$ amino-acids and $m$  is the electron effective mass, here taken the same of the bare value.
\par
Figure \ref{fig:2} reports the evolution of the simulated current responses for two different voltage values, 1 V corresponding to DT regime, and 5 V corresponding to FN regime.
We can observe the increasing number of spikes when passing from the low to the high voltage. 
At 5 V the number of spikes is so high that it becomes difficult to classify them as "extreme events". 
On the other hand, by lowering the bias value, the number of spikes lowers but does not definitely disappear, as shown in the same figure for the case of 1 V.
\par
Figure~(\ref{fig:3}) reports the variance of current fluctuations corresponding to the I-V characteristic of  Fig. (\ref{fig:1}).
Here we notice a rather abrupt increase of the  variance of current fluctuations in concomitance with the cross-over region of the I-V characteristic.
The giant  increase, for about five orders in the magnitude of current variance, is associated with the opening of low resistance paths between contacts, originated by the establishing of the FN tunneling regime.
\par
Figure {\ref{fig:4}} reports the probability distribution functions  (PDFs) of current fluctuations at different voltages covering the cross-over region of the two tunneling regimes. The variables have been rescaled
as suggested in \cite{Bramwell,PhysicaA,Bertin} to evidence the
"universal" behaviour of all the curves.
In the inset of Fig. (\ref{fig:4}) the PDFs are shown before rescaling. 
The divergence from Gaussian behaviour is not completely surprising, being partially due to the small size of the system under consideration \cite{PhysicaA}, but also to the critical condition of a system in which, for all the considered voltages, two different tunneling regimes are in competition. 
On the other hand, this kind of systems, and their fluctuations in particular, have received in the last years  a relevant attention, mainly due to the possibility to trace back them to an unique root. 
The amount of results of recent past \cite{Bramwell,Noullez,Aji,Bertin,Brey,Clusel,Ciliberto,Manzato} suggests this behaviour should be described by a generalized Gumbel distribution, one of the reference distributions for the statistical modeling of extreme events \cite{coles05}. 
\section{The scaled Gumbel distribution}
\begin{figure}[htb]
\centering\noindent
\includegraphics[scale=0.7]{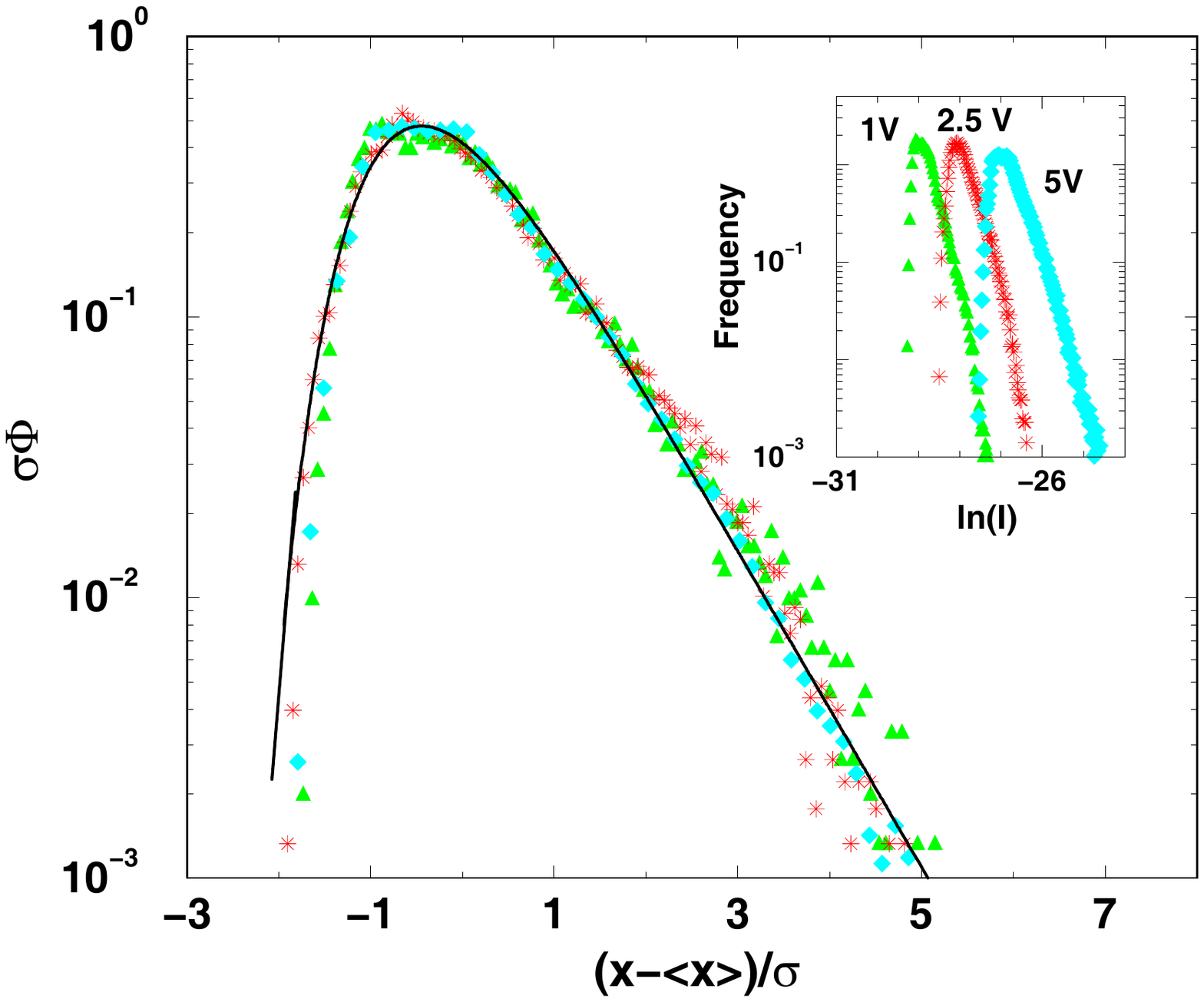}
\caption{Scaled PDFs of current fluctuations for different bias values. Symbols refer to simulations, continuous line to the scaled Gumbel distribution of order 1 (see Eq. 5 in text). 
The inset reports the PDFs  at the given bias before applying the scaling.}
\label{fig:4}
\end{figure}

The results reported in the inset of Fig. (\ref{fig:4}) are found to exhibit a skew shape, resembling the popular, almost ubiquitous PDF studied by Bramwell-Holdsworth-Pintor (BHP) \cite{Bramwell}. 
This distribution was found in very different contests going from fluidynamics, to self-organized critical systems, and to resistance fluctuations \cite{Bramwell,Clusel,PhysicaA}. 
Therefore, we analyze current fluctuations to understand whether  BHP can be, also in the present case, an adequate PDF.
%
To this purpose, data are scaled by using the mean value and the variance in the conventional way\cite{Bramwell}:
\begin{equation}
z \rightarrow \frac{(z-<z>)}{\sigma} \equiv x, \qquad\qquad  \Pi(x)\rightarrow \sigma\Pi \equiv G(a,x)
\end{equation} 
with $\Pi(x),\ <x>, \ \sigma$ assuming the standard meaning of the PDF, the mean value of the variable and the standard deviation, respectively, and with  $z=ln(I)$. The numerical parameter $a$ is related to the specific system conditions
\cite{Bertin}.
\par
The choice of the logarithm of current instead of the current itself is suggested by the slow dependence of the former on the applied voltage \cite{Bramwell,Sylos}.
As illustrated by Fig. (\ref{fig:4}), after rescaling, PDFs related to different
bias values ($1 \div 5$ V) collapse in a single one, say $G(1,x)$ which, on the other hand, is not the BHP but the generalized Gumbel function of order 1:
\begin{equation} 
G(1,x)=  \theta(a)\,e^{-(\theta(a)x+\gamma)-e^{-(\theta(a)x+\gamma)}}
\label{eq:g1}
\end{equation}
with $a=1$, $\theta^{2}(a)$ the trigamma function and $\gamma$ the Euler constant. This PDF has been also called the \textit{scaled} Gumbel distribution function, and has been found of interest in describing the fluctuations of the conditional galaxy density \cite{Sylos} and is a special case of the PDF initially introduced by Clusel-Fortin-Holdsworth \cite{Clusel} (CFH). 
For the sake of completeness, the CFH distribution recovers the BHP when $a=\pi/2$.   
\begin{figure}[htb]
\centering\noindent
\includegraphics[scale=0.7]{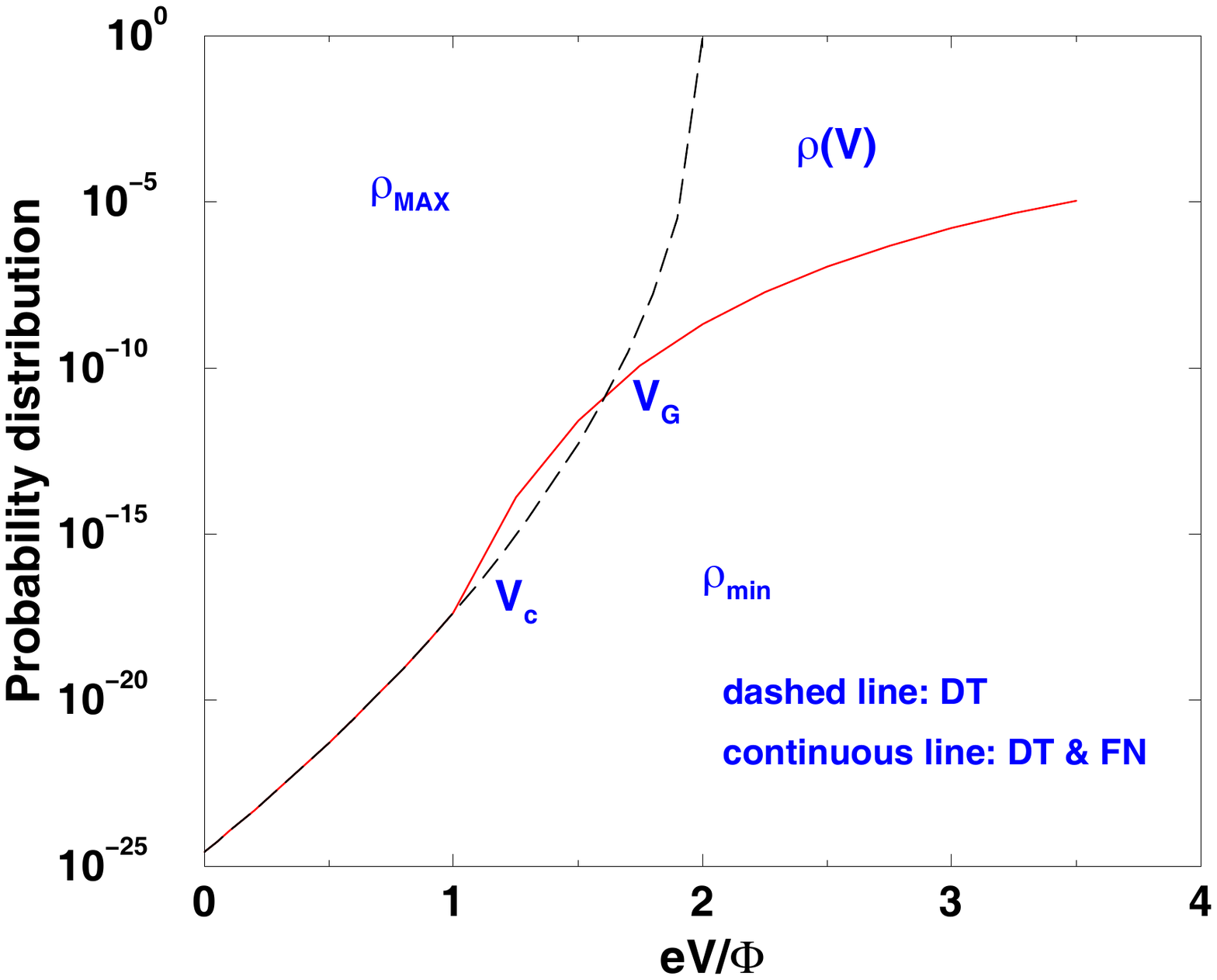}
\caption{Tunneling transparency given by Eqs. (2) and (3), here interpreted as probability distribution of the the resistivities pertaining to the two tunneling phases, vs normalized voltage.}
\label{fig:5}
\end{figure}
The CFH distribution has been found in many systems in which at least one phase transition appears \cite{Bertin,Ciliberto} and usually, as made for BHP, its origin was related to the existence of a correlation length larger than the system size. 
On the other hand, it is the natural generalization of the Gumbel distribution of extreme events \cite{Gumbel}; therefore the question of how the two interpretations can converge remains an open one. 
The question is discussed in Ref. \cite{Bertin}, and the CHF distribution with integer values of $a$ is described as the  distribution of the $a$-th larger/smaller value of the variable,
here identified in the high and low resistivity values.
The analogy with natural (for example the climate) extreme events is evident: they are considered extreme or exotic under standard  environmental conditions, otherwise when environmental conditions change in a significative way, they  becomes the most probable response \cite{Katz}.   
\section{Results and discussion}
The INPA model was used in the  "large tip" version that involves several amino acids in the contact between the protein and the tip. 
In such a way, the real extent of the AFM tip (transversal size of about 200 nm) can be taken into account. 
The single protein results fit the sample response (see Fig. 1) when the resistivities to be associated with the links are taken as $\rho_{MAX}=8  \times 10^{13} \ \Omega$ \AA, $\rho_{min}=4 \times 10^{6} \ \Omega$ \AA, which correspond to a rough estimation of $N \approx 10^{9} $ trimers in
an area of about $2\times 10^{11} nm^2$ \cite{PRE11}. 
The energy-barrier height is $\Phi=219 \ meV$.
We remark that after the proper scaling, the PDFs of current fluctuations
shown in Figs. (2) and (3) converge to the single $G(1,x)$ curves, as reported in Fig. (\ref{fig:4})
\par
The sharp cross-over between the DT and FN tunneling regimes admits for a phase transition interpretation, with DT and FN being the two phases.
To this purpose, Fig. (\ref{fig:5}) reports the tunneling transparency used to model the sequential tunneling mechanism (see Eqs. (2) and (3) ), here called probability distribution,  vs the applied voltage.
The cross-over between the DT and FN regimes occurs in the region where the two curves (dashed and continuous lines) cross each other, respectively at the two voltages: the critical voltage, $V_c$, and, with the same meaning of the Ginzburg temperature, \cite{Antunes}, the Ginzburg voltage, $V_G$. 
For voltages in the region $V < V_c$, the resistivity of each link can take only the two values $\rho_{MAX}$ and  $\rho_{min}$, respectively.
This is the case of a squared shape for the tunneling barrier. 
For voltages $V \geq V_c$, which corresponds to a triangular shape of the tunneling barrier, the single value of $\rho_{MAX}$  smooths to a continuous range of $\rho(V)$.
Accordingly, we conjecture that the region of bias inside the values $V_c \div V_G$ corresponds to the consolidation of the FN tunneling regime, the
new phase for the charge transport through the physical system.  
In other words, by approaching the voltage region $V_c \div V_G$, for $V_{i,j}$ in the DT regime,  each link will choose a resistivity values between $\rho_{MAX}$ and $\rho_{min}$. 
By contrast,  for $V_{i,j}$ in the FN regime,  each link will choose a resistivity values between $\rho(V)$ and $\rho_{min}$.
For voltages above the value of $V_G$, the FN phase will take place for all the links. 
\section{Conclusion} 
The paper reports a microscopic model of current-voltage and current fluctuations in a two terminal sample of nanometric width where the active region is provided by a monolayer of bacteriorhodopsin. 
In particular, charge transport has been analyzed in a very wide range of voltages up to  7 V.
Here current response exhibits a strong superlinear behavior at increasing voltage, typical of a transport controlled by a tunneling mechanism.
A unifying microscopic description, the so-called INPA model, is able to interpret the experiments on the basis of two tunneling regimes, a direct one (DT) at low bias and an injection one (FN) at high bias.
Both tunneling regimes are assumed of sequential type and they occur between neighbouring amino-acids.
Within this model, we obtain good agreement between theory and I-V experimental characteristics. 
More interesting, current fluctuations are found to exhibit a sharp increase for about five order of magnitude in correspondence of the cross-over between the  DT and FN tunneling regimes.
The associated probability distribution function is found to follow a universal behaviour that is characterized by a generalized Gumbel distribution, which is typical of  systems close to a phase transition.
In the present case, the phase transition is associated with the cross-over between two tunneling regimes each of them taken as a single phase, and a kind of phase diagram is individuated in the voltage dependence of the tunneling transparency that has been used to interpolate the two tunneling regimes.

\section*{Acknowledgments}
This research is supported by the European Commission under the Bioelectronic Olfactory Neuron Device (BOND) project within the grant agreement number 228685-2.


\begin{thebibliography}{99}
%
\bibitem{Gumbel}
E.J. Gumbel, \Journal{J. Am. Stat. Ass.}{55}{1960}{698}.
\bibitem{Bramwell}
S.T. Bramwell, P.C.W. Holdsworth, and J.-F. Pintor,
\Journal{Nature}{396}{1998}{552};
S. T. Bramwell, K. Christensen, J.-Y. Fortin, P. C. W. Holdsworth, H. J. Jensen, S. Lise, J. M. L{\'o}pez, M. Nicodemi, J.-F. Pinton, and M. Sellitto
\Journal{\PRL}{84}{2000}{3744}.
%
\bibitem{Noullez}
A. Noullez and J.-F. Pinton,
\Journal{Eur. Phys.J. B}{28}{2002}{231}.
%
\bibitem{Aji}
V. Aji and N. Goldenfeld \Journal{\PRL}{86}{2001}{1007}.
\bibitem{Bertin}
E.~Bertin,
\Journal{\PRL}{95}{2005}{170601}.
\bibitem{Brey}
J. Javier Brey, M. I. Garcı´a de Soria, P. Maynar, and M. J. Ruiz-Montero, \Journal{\PRL}{94}{2005}{098001}
%
\bibitem{Clusel}
M. Clusel, J.Y. Pintor, and P.C.W. Holdsworth \Journal{\EPL}{76}{2006}{1008}.
\bibitem{Ciliberto}
S. Joubaud, A. Petrosyan, S. Ciliberto, and B. Garnier, \Journal{\PRL}{100}{2008}{180601}.
\bibitem{Manzato}
C. Manzato, A. Shekhawat, Phani K.V.V. Nukala, M.J. Alava, J.P. Sethna, and Stefano Zapperi, \Journal{\PRL}{108}{2012} {065504}
%
\bibitem{Corcelli}
A. Corcelli, M. Coletta, G. Mascolo, F.P. Fanizzi, and M. Kates,
\Journal{Biochemistry}{39}{2000}{3318}.
%
\bibitem{Epl} E. Alfinito and L. Reggiani, \Journal{Europhys. Lett.}{85}{2009}{86002}.
%
\bibitem{PRE11}
E. Alfinito, J.-F. Millithaler, and L. Reggiani, 
\Journal{\PRE}{83}{2011}{042902};
E. Alfinito, and L. Reggiani, 
\Journal{\PRE}{81}{2010}{032902}.
%
\bibitem{Gomila}
I. Casuso, L. Fumagalli, J. Samitier, E. Padr{\'o}s, L. Reggiani, V. Akimov
and G. Gomila, 
\Journal{\PRE}{76}{2007}{041919}.
%
\bibitem{Nano}
E. Alfinito, C. Pennetta, and L. Reggiani, \Journal{Nanotechnology}{19}{2008}{065202}.
%
\bibitem{Wang}
W. Wang, T. Lee, and M.A. Reed, \Journal{Rep. Prog. Phys.}{68}{2005}{523}.
 %
 \bibitem{Simmons}
J.G. Simmons,
\Journal{\JAP}{34} {1963}{1793}.
%
\bibitem{PhysicaA}
C. Pennetta, E. Alfinito, L. Reggiani, S. Ruffo, \Journal{Physica A}{340}{2004}{380}.
\bibitem{coles05}
H.J. Coles,  and  M.N. Pivnenko,  
\Journal{Nature}{436}{2005}{997}
\bibitem{Sylos}
T. Antal, F. Sylos Labini, N.L. Vasilyev, Y.B. Baryshev, 
\Journal{Eur.Phys. J.}{88}{2009}{59001}.
%
%

%
\bibitem{Katz}
R.W. Katz, \Journal{Climatic Change}{100}{2010}{71}.
\bibitem{Antunes}
N.D. Antunes, L.M.A. Bettencourt, and W.H. Zurek, \Journal{\PRL}{82}{2000}{20824};
 E. Alfinito, and G.Vitello\Journal{\PRB}{65}{2002}{054105}.
\end{thebibliography}
\end{document}